# ADAPTIVITY AND PERSONALIZATION APPLICATION SCENARIOS IN EPARTICIPATION

*Babis Magoutas[1], Gregoris Mentzas[2]*

*Adaptivity and personalization technologies appear not to be very much used in e-participation projects to date. These technologies are commonly used to overcome the overflow of information and service providers adopt them in order to acquire a better knowledge of their end-users and optimize their service offerings. In this paper we investigate the potential of adaptivity and personalization principles and technologies when applied to the eParticipation field and more specifically, to eParticipation websites. Potential application scenarios of these technologies in the context of policy engagement and active participation of citizens in democratic decision-making are defined and their impact in eParticipation is examined.*

## 1. Introduction

As the World Wide Web matures, it makes leaps forward in both size and complexity. In this expanding environment, the needs and interests of individual users become buried under the sheer weight of possible viewing choices. Beside the problem of finding interesting information on the web, users experience big problems in navigating through the content of a web portal, especially regarding the efficiency of that process.

One major problem is the lost-in-hyperspace one: In the rich link structure of a portal, users can easily get overwhelmed and become unable to navigate effectively. There is a growing body of empirical evidence to suggest that users tend to make poor decisions in traditional systems as the navigational freedom given to the user leads to comprehension and orientation difficulties in the sense that users may become spatially disoriented, lose sight of objectives, skip important content, choose not to answer questions, look for stimulating rather than informative material or simply use the navigational features unwisely [13].

The second major problem is the one-size-fits-all: Websites provide (from the user's point of view) relatively static content and are viewed by diverse users. This may cause difficulty for those who have less background; also it can be redundant for those who already know the information; and be uninteresting for others. In other words, since the user population is relatively diverse, traditional static applications suffer from an inability to satisfy the heterogeneous needs of the many users [13].

In order to cope with these problems, there is a demand for intelligent tools and structures which can make navigation of the sites easier for users and maximize the quality and completeness of their experience. To counter this, there has been a rise in research in adaptive websites and personalization, a combination of data mining, machine learning, user modeling,

---

[1] National Technical University of Athens, 157 80, Zografou, 9 Iroon Polytechniou str., elbabmag@mail.ntua.gr
[2] National Technical University of Athens, 157 80, Zografou, 9 Iroon Polytechniou str., gmentzas@mail.ntua.gr

Human Computer Interaction (HCI), optimization theory and graph theory which seeks to sift through the tides of possible information and provide users with a high-quality stream of information.

In respect to e-participation, where a number of citizens with different needs and expectations is addressed, adaptivity and personalization can help to structure the complex area thereby creating a new way in which citizens engage in the discourse with politicians and governments. To our knowledge applications in the convergence between personalization and e-participation are limited to date. In this paper we investigate the potential of adaptivity and personalization principles and technologies when applied to the eParticipation field and more specifically, to eParticipation websites. Potential application scenarios of these technologies in the context of policy engagement and active participation of citizens in democratic decision-making are defined on the conceptual level for various participation areas and their impact in eParticipation is examined.

It should be noted that our primary goal is not to present a systematic review of the state of the art of adaptivity and personalization, neither to provide a categorization scheme for the various participation areas, but rather to identify future application scenarios of these technologies for the various participation areas, in order to address the aforementioned convergence gap.

The paper is structured as follows. After this brief introduction, in section 2 we set the context and give the theoretical background by providing an overview of adaptation targets and techniques used for personalizing the user experience. After providing a comprehensive understanding of the scope of research, we identify in section 3 potential future application scenarios of adaptivity and personalization for the following participation areas: information provision, consultation and deliberation, community building, mediation, spatial planning and campaigning. Finally in section 4 we sketch some conclusions by discussing the two-fold impact of adaptivity and personalized technologies in e-participation.

## 2. Theoretical Background

### 2.1 Definitions

Adaptivity is a particular functionality that alleviates navigational difficulties by distinguishing between interactions of different users within the information space. Adaptive Systems employ adaptivity by manipulating the link structure or by altering the presentation of information, based on a basis of a dynamic understanding of the individual user, represented in a user model [13]. An adaptive hypermedia system is a hypermedia system which reflects some features of the user in the user model and apply this model to adapt various visible and functional aspects of the system to the user [10], [4]. A system can be classified as an Adaptive Hypermedia one if it is based on hypermedia, has an explicit user-model representing certain characteristics of the user, has a domain model which is a set of relationships between knowledge elements in the information space, and is capable of modifying some visible or functional part of the system based on the information maintained in the user-model [10], [4], [2]. The broadest definition of an adaptive website is a website which changes based on the way it is used [18]. Changes can take on many forms, as they may either be immediate (as in the case of recommendation systems) or gradual (as in the case of systems which suggest changes to a website administrator).

A relatively new research area, very closely related with adaptive web systems is web personalization. Web personalization has a more extended scope than adaptive hypermedia, exploring adaptive content selection and adaptive recommendation based on modeling user interests and it is primarily used in the e-business application domain [13]. As is often the case with a good marketing buzzword, the term personalization is used rather loosely [5]. It has come to stand for an ultimate goal of customer relationship management by businesses, supporting for example one-to-one marketing. It has also come to mean delivery of information of high relevance to an individual, in the context of receiving from a large body of information only the part that is of interest to an individual or a group of individuals [30]. In [30] personalization is defined as delivering to a group of individuals relevant information that is retrieved, transformed, and/or deduced from information sources, while authors of [21] state that, web personalization refers to the whole process of collecting, preprocessing, classifying and analyzing Web data, and determining based on these the actions that should be performed so that the user is presented with personalized information. In [20] personalization is defined as the adaptation of products and services by the producer for the consumer using information that has been inferred from the consumer's behavior or transactions, by using technology.

In summary, personalization takes place between one or several "providers" of personalized "offerings" and one or several "consumers". Personalized "offerings" include content (such as Web pages and links), product and service recommendations (such as for books, CDs, and travel packages), e-mail, information searches, dynamic prices, and products for individual consumers (such as custom CDs).

## 2.2 Adaptation Targets and Techniques

There are two major questions that must be taken into account when an adaptive/personalized system or application is considered.

- What can be adapted, meaning which are the targets for adaptation
- Which techniques are used by the system in order to collect user information and create the user profile, which is subsequently used to adapt to the user.

In this section we describe the targets of adaptation, as well as the adaptation techniques that are commonly found in adaptation systems.

### 2.2.1 Targets for Adaptation

The heart of an adaptive web system is its ability to change in response to the way it is used. This section provides an overview of the kinds of changes that such a system may perform. It should be recognized that the content, presentation and links of a web page are closely related, so there is bound to be crossover between these categories.

One of the basic modifications that might be made is to change the **content** of the web page, based on the model that the system has been able to deduce about the user [17]. Content might be added to or removed, or it might be simply rearranged [6]. These modifications might be done to accomplish several things, including optional explanations/detail and opportunistic hints [17], substitution of content [12] and personalized recommendations [17].

In addition to modifying the content of the page, one can also change the way it is presented in order to serve a user. Most of the research on adaptive **presentation** deals with adaptive text presentation, and mostly with canned text presentation. Examples of adaptive canned text presentation include inserting or removing fragments, dimming fragments and fragment coloring [7], [1].

Adaptation of **navigation** realizes adaptation by changing the links of the system [17]. This adaptation speeds up the search for a particular page and helps to avoid the problem of users lost in hyperspace. Examples of this type of adaptation include direct guidance, adaptive link sorting, adaptive link hiding guidance, adaptive link annotation, adaptive link generation and map adaptation [3].

It is also possible for an adaptive system to modify the long-term **structure** of the website in a "permanent" fashion, rather than the per-request temporary fashion suggested above. Usually, the final decision to add or remove a page or atom should be ultimately made by some human administrator, but the indication of whether it should be added or dropped can be made by the system. Several indications may be given by the system, including for example new index pages and permanent new link suggestions [24].

*2.2.2 Adaptation Techniques*

The techniques available to collect information about users and the methods used to process such information to create user profiles and provide adapted content, presentation and/or structure, are varied. Most web personalization techniques fall into four major categories: content-based filtering, collaborative filtering, rule-based filtering and web usage mining.

Content-based filtering systems recommend items to users (such as content, services, and products) like the ones they preferred in the past. Content-based methods analyze the common features among the items a user has already rated highly. Only the items similar to user's past preferences are then recommended. In other words these systems are solely based on individual users' preferences [23], as they use correlations between the content of the items and the user's preferences in order to build the user model and adapt to the individual user. All of the content-based approaches represent items by the "important" words in the items.

Collaborative filtering systems invite users to rate the objects or divulge their preferences and interests and then return information that is predicted to be of interest to them. These systems make automatic predictions (filtering) about the interests of a user by collecting preferences from many users (collaborating) and then recommend items to the user that people with similar tastes and preferences have liked in the past. The basic idea underlying collaborative systems is that the adaptation is based on the experiences of a population of users, rather than on an individual user profile. This is based on the assumption that users with similar behavior (e.g., users that rate similar objects) have analogous interests and that those who agreed in the past tend to agree again in the future [26].

Rules-based personalization is the delivery of personalized content based on the subjection of a user's profile to set rules or assumptions [8]. The rules are used to affect the content served to a particular user, based on relationship analysis. Rules-based personalization systems use business logic embedded in conditional (if/then) statements to create content display. Under rules-based personalization, a user's known preferences fulfill certain criteria, and corresponding content is served accordingly. A system administrator typically uses a visual interface to input if/then criteria, specifying each condition and the content which should be recommended in response.

Web usage mining techniques rely on the application of statistical and data mining methods (e.g. association rule mining, sequential pattern discovery, clustering, and classification) to the web log data, resulting in a set of useful patterns that indicate users' navigational behavior. These patterns are used in order to predict user behavior and provide personalized experience while users interact with the Web [28].

## 3. Application Scenarios in eParticipation

Adaptivity and personalization technologies appear not to be very much used in e-participation projects to date. Nevertheless, by analysing the "Participation" areas described in [9], we will provide some possible application scenarios of adaptivity and personalization technologies in e-participation and we will show the significant potential that these technologies can possess to the engagement of citizens, to the strengthen of citizenship and public-involvement at the national, regional, and local levels. Some of the participation areas defined in [9], in which adaptivity and personalization could come into consideration, are shown in Table I:

| e-Participation Area | Description |
|---|---|
| Information Provision | ICT to structure, represent and manage information in participation contexts |
| Consultation | ICT in official initiatives by public or private agencies to allow stakeholders to contribute their opinion, either privately or publicly, on specific issues |
| Deliberation | ICT to support virtual, small and large-group discussions, allowing reflection and consideration of issues |
| Community Building | ICT to support individuals come together to form communities, to progress shared agendas and to shape and empower such communities. |
| Mediation | ICT to resolve disputes or conflicts in an online context |
| Spatial Planning | ICT in urban planning and environmental assessment |
| Campaigning | ICT in protest, lobbying, petitioning and other forms of collective action (except of election campaigns) |

*Table I: Participation Areas Relevant to Adaptivity and Personalization*

It should be noted that the above list is by no means exhaustive. We have selected the above-mentioned areas mainly based on work reported in [9]; however, we acknowledge that other participation areas where adaptivity and personalization can be applied exist and furthermore different descriptions of the listed areas might be used. Nevertheless we do believe that we cover the most important ones and that the application scenarios described in the rest of this section can be easily mapped to different descriptions of the various participation areas.

### 3.1 Information Provision

The information provision participation area is strongly related with the first level of e-participation (e-informing) [16]. E-informing aims at providing the public with balanced and objective information to assist them in understanding the problem, alternatives, opportunities, solutions as well as in gaining insights of the political systems and the deliberations. Major stakeholders need access to background information and issues on the political agenda, like technical and legalistic information, in order to be able to discuss and participate. This is more relevant to participation exercises that are close to the draft policy stage, as higher demands on citizens' ability to understand technical and legalistic statements exist in this stage [22].

The necessity to fairly deal with different categories of citizens - belonging to both sides of the digital divide, and having different experience and knowledge - claim for the personalization of the information supplied. For that reason the process of making background information suitable for the target audience is a high priority issue. We have identified three possible e-participation application scenarios of adaptivity and personalization, which are related with information provision.

The first application scenario concerns the provision of personalized views of background downloadable documents based on user profiles. This approach has been applied in [14] for providing personalized e-government services. The main idea underlying this research is that some legal documents or some of their parts have a limited applicability only to specific classes of citizens. For example, some articles of a tax-related legal document are applicable only to unemployed persons, some to self-employed persons, some others to public servants only and so on. Ideas from this e-government application of adaptivity and personalization technologies can be incorporated in e-participation scenarios as well. For example legal documents and norms can be adapted and each stakeholder can view versions of these documents containing only articles which are interesting for him/her. This idea can be also adopted for the personalization of other types of documents, like technical documents concerning specific deliberation issues or documents describing the policy formation processes.

The second application scenario concerns adaptive and personalized features of the e-participation site that enable the accessibility and understand-ability of the information presented. User's knowledge, expertise, backgrounds and disabilities can be taken into account in order to personalize the presentation of information:

- Information in different levels of detail can be provided to each citizen according to his knowledge about the topic presented, by making use of text fragments adaptivity.

- Help and support information, about the use of discussion forums should be more detailed for users with little knowledge, while experience users should not be bothered with irrelevant information.

- Linguistic versions of the e-participation site can be adapted to individuals depending on their language understand-ability.

- The font size can be varied based on citizens' vision ability.

Such features will encourage full participation of people with disabilities and will bridge the digital divide by ensuring inclusiveness across a diversity of needs. Furthermore such a use of personalized features and services in e-participation sites facilitates the interaction of citizens with the government, hence improving the access to government as well as the satisfaction of citizens. This has as consequence that citizens are more trusting of the e-participation web site, which in turn leads to loyalty. With this way personalization features foster greater citizen interest and involvement in public issues and indirectly drive citizens to engagement with many e-participation features, like e-deliberation, e-petitioning, etc.

A third application scenario is related with the emerging technology of webcasting. Taking into account the emerging trend that a lot of users prefer to view rather than read information, multimedia tools such as webcasting can be used by governmental authorities for communicating key messages and involving the public. The role of webcasting in e-participation has been investigated in the eTEN EU funded project eParticipate [11]. We believe that interaction with the public using webcasting would be enhanced if personalization technologies are considered. More specifically two levels of personalization have been identified. In the first level the most relevant and interesting streams could be recommended to each citizen, while personalized streams to each citizen, according to his user profile, can be provided as the second level of personalization.

### 3.2 Consultation, Deliberation

One of the major activities of consultation and deliberation participation areas, include the expression of citizens opinions on specific themes and issues of the political agenda that are affecting them, through various means such as discussions, surveys and polls. Themes and issues of deliberation and consultation are proposed either by public authorities or by citizens. By applying personalization and adaptivity features in the aforementioned participation areas, the themes/issues of deliberation or consultation can be personalized according to citizens' interests, in terms of recommendations of interesting topics/themes to each individual citizen.

This application of personalization is very important especially in large-scale e-participation sites where the overload of deliberation themes could possible discourage citizens' engagement. In this possible application, attention should be given to inclusion of every citizen that wants to participate in the process. In other words, all possible issues and themes that are subject to consultation and deliberation should be visible to everyone, i.e. recommendations should be viewed as a means to increase the timeliness and usability of the process and not as a way to filter out specific citizens' categories.

Adaptive surveys are another possible application of the technologies under discussion. While surveys can be used in many participation areas, the participation area that is mainly supported by surveys is "consultation", as described in [9]. Surveys are mainly used to research views, attitudes and experiences of participants and are realized as web-based questionnaires, where the e-participation website shows a list of questions, which users answer and submit their responses online. A common problem with web-based surveys is that users are usually reluctant to participate, especially when the time needed to complete the questionnaire, or the number of questions, increase [15].

This problem can be addressed if the list of questions to be given to each participant is not fixed, but composed dynamically from a predefined set of questions, or if the list of alternative answers for a single question is adapted based on some criteria. This approach, which has been realized for measuring the quality of e-government service in [19], can be also incorporated in the context of consultation e-participation area. Criteria for adaptation may include in this case, user profile data such as participant's interests, geographical information, experience and abilities. With this way not only the response rate increases, but also citizens' feedback is targeted, as each citizen is providing feedback for the consultation topics/questions of the questionnaire that he/she is more knowledgeable and interested. In other words citizens with insight and knowledge in a given topic are getting more in touch with decision makers.

### 3.3 Community Building

Forums can play an important role in public dialogue. Communities are built around a variety of topics and citizens participate by contributing ideas that could be very useful for authorities in order to design their policies and make their decisions. Topics or threads of messages, in discussion forums can be recommended to each citizen, in a personalized way, according to his/her previous engagement in e-participation exercises. For example citizens that have participated heavily in threads where privacy issues in traditional offline advertising are discussed, can be invited to participate in discussions about privacy issues in online information systems, as well. Such invitations can be provided either in terms of recommended links or by using the information push communication paradigm. In the latter case citizens can receive updates and alerts (e.g. newsletters, emails) about new threads concerning a topic of their interest.

### 3.4 Mediation

Mediation is a form of dispute resolution aiming to assist disputants in reaching an agreement. As adaptivity and personalization techniques provide a personalized view of a domain, based on each participant's experience and expertise, their use could be applied for conflict prevention by unveiling concepts and relations whose ambiguity may lead to misunderstanding or dispute or conflict. More specifically advanced and experienced participants should be provided with a coarse-grained detail about concepts and relations, in contrast to participants that need more explanations and disambiguation, where a fine-grained detail level should be given. With this way a better understanding of concepts and relations could be achieved by all participants and dispute and conflict would be mitigated.

### 3.5 Spatial Planning

A possible application scenario of personalization for national e-participation sites is the personalization of engagement concerning spatial planning, based on geographical profiles of participants. Although this application scenario could be relevant for e-participation sites of an urban, municipal or prefecture scale, national scale e-participation sites that spread large geographical areas are more appropriate. According to this scenario the e-participation site can take into account geographical and location-based data about participants – like their cities, villages, municipalities, geographical areas of interest or even their position in a mobile context - in order to provide a personalized view of issues and themes that are subject to citizens' engagement. This allows the filtering of irrelevant data, and thus the engagement in the process of spatial planning in made more efficient and usable.

An important risk of opening the spatial planning process to the public, is that some private own corporations that do business in the real estate field, will probably try to affect the political agenda or the final decision, in order to increase their profit. By weighting participants' opinions based on their geographical profile, the aforementioned conflict of interest can be mitigated, as the weights that will be attributed to those directly affected by the project, (e.g. habitants or people with strong relations with a specific geographic area) will be bigger.

### 3.6 Campaigning

The campaigning participation area refers to an organized effort to influence the decision making process. E-petitioning and e-protesting contribute to such an effort. Online petitions/protests allow citizens to sign in for a petition/protest by adding their name and address online. This participation area could be enhanced if personalization and adaptivity technologies are considered. The number and topic of petitions/protests that a citizen has participated can be used in order to build a citizen profile, which encapsulates his/her political interests. This profile can then be used by personalization techniques, forming the basis for inviting the citizen to participate in other petitions/protests of similar content.

Another important possible application of adaptivity and personalization to the campaigning participation area is related to the adaptation of the back office. Handling of citizens' petitions/protests can be assigned to the most appropriate governmental authority, or to the most appropriate public servant, depending on their knowledge, availability and role. A similar approach has been adopted by the EU-funded IST project FIT [27], where flexibility and adaptivity of the back office has been increased by extending process models by business rules.

### 3.7 Summary

Table II summarizes the identified potential application scenarios of adaptivity and personalization for the various e-participation areas.

| e-Participation Area | Application Scenario |
|---|---|
| Information Provision | - Personalized views of background downloadable documents<br>- Adaptive/personalized features of the e-participation site enabling information accessibility and understand-ability<br>- Recommendation of interesting webcast streams and personalization of webcast streams |
| Consultation, Deliberation | - Recommendation of interesting consultation/ deliberation issues and themes<br>- Adaptive consultation surveys |
| Community Building | - Content-based recommendations of forum topics and threads |
| Mediation | - Personalized views of the domain of dispute and conflict, based on each participant's experience and expertise |
| Spatial Planning | - Location-based recommendation of spatial planning issues |
| Campaigning | - Content-based filtering of e-petition and e-protest issues<br>- Back office adaptation: Handling of citizens' petition and protests by the most appropriate authority |

*Table II: Potential Application Scenarios of Adaptivity and Personalization in various e-Participation Areas*

## 4. Conclusions

Adaptive and personalized systems are used in several application areas where the hyperspace is reasonably large and where a hypermedia application is expected to be used by users with different goals, knowledge and backgrounds. E-participation has the characteristics described above, thus it is an application area where personalization and adaptation techniques and systems can be applied.

The impact of adaptivity and personalized technologies in e-participation is two-fold, having direct as well as indirect results. Direct impacts have mainly a qualitative nature while the nature of indirect impacts is mainly quantitative. On the one hand direct impacts are related with the increase of the quality and completeness of citizens' engagement in the discourse with politicians and governments, as the application of personalized techniques offers higher functionality and an optimized usability to e-participation exercises (which are otherwise mutually exclusive attributes).

On the other hand, the provision of personalized support and services to citizens has as prerequisite the availability of deep knowledge about each individual. In this way by using prior interactions and knowledge with/of citizens, long-term and more stable relationships with them can be built. These stable relationships, as well as the increase of citizens' satisfaction attributed to their personalized experience, can leverage the enlargement of citizens' participation though nationals or municipals e-participation sites.

Although adaptivity and personalization offer opportunities to make the engagement of citizens at public issues more effective and efficient, some drawbacks and limitations should be considered, when applying these technologies to e-participation. In contrast to private organizations, governments have to offer their services to each citizen and treat each one on an equal basis. Therefore personalization and adaptation functionalities should be optional and should be viewed by public organizations as a means to increase user usefulness and ease of use and not as a way to filter out specific citizens' categories.

Another important limitation of applying adaptivity and personalization technologies to e-participation is related to user privacy and data protection issues. Improper acquisition, use, storage and control of personal information and privacy invasion are some of the most important privacy threats and concerns [29]. Privacy infringement issues make it difficult for governments to adopt personalization strategies from the commercial sector, while many agencies may in fact be restricted from engaging in personalization by various privacy acts [25]. As a minimum response to privacy concerns, the public organizations should declare in a clear and concise manner a privacy statement describing the kinds of information gathered and the policies for using and sharing personal information.

In this paper the applicability of adaptivity and personalization technologies to the area of e-participation has been discussed in the context of possible future application scenarios. E-participation programs can be applied in various levels of public administration, such as in a national or municipal level. The higher the level of public administration that the e-participation program is applied, the more applicable the technologies of adaptivity and personalization are. This is true because of the increasing need to address the "one size fits all" problem in a setting of increased availability of information. In other words the information overload, which is the main problem that the technologies under discussion address, is bigger in national than in municipal level. On the other hand e-participation is an emerging area of research and it has been mainly applied in pilot environments at the municipal level of administration. Considering these, we can conclude that adaptivity and personalization technologies have a very big potential for future large scale e-participation programs.

## 5. Acknowledgments

Research reported in this paper has been partially financed by the European Commission in the project DEMO-net (The Democracy Network) under the IST FP6 programme.

## References

[1] Ardissono L., Console L., Torre I., An adaptive system for the personalized access to news. In AI Communications, 14(3):129–147, 2001

[2] Brusilovsky P., Nejdl W., Adaptive Hypermedia and Adaptive Web. © 2004 CSC Press LLC, 2004.

[3] Brusilovsky P., Adaptive navigation support. In: Brusilovsky, P., Kobsa, A., Neidl, W. (eds.): The Adaptive Web: Methods and Strategies of Web Personalization. LNCS, Vol. 4321. Springer-Verlag, Berlin Heidelberg New York, 263-290, 2007

[4] Brusilovsky P., Adaptive Hypermedia. User Modeling and User-Adapted Interaction, 11: 87-110, 2001.

[5] Communications of the ACM, Special Issue on Personalization, 2000.

[6] De Bra P., Aroyo L., Chepegin V., The next big thing: Adaptive web-based systems. Journal of Digital Information, 5(1), 2004.

[7] De Bra, P., Stash N., Smits D., Creating Adaptive Web-Based Applications, Tutorial at the 10th International Conference on User Modeling, Edinburgh, Scotland, 2005.

[8] Deitel H.M., Deitel P.J., Steinbuhler K., e-Business and e-commerce for managers. Prentice Hall 2004.

[9] DEMO-net Consortium (Macintosh A. (Ed.)), D5.1 – Report on current ICTs to enable Participation, 2006.

[10] Eklund J., Sinclair K., An empirical appraisal of the effectiveness of adaptive interfaces of instructional systems. Educational Technology and Society 3 (4), ISSN 1436-4522, 2000.

[11] eParticipate eTEN project home page, Available at http://www.eparticipate.org/Project_overview.htm

[12] Germanakos P., Mourlas C., Adaptation and Personalization of Web-based Multimedia Content. In Multimedia Transcoding in Mobile and Wireless Networks, 2008.

[13] Germanakos P., Mourlas C., Isaia C., Samaras G., An Optimized Review of Adaptive Hypermedia and Web Personalization - Sharing the Same Objective. Proceedings of WPRSIUI 2005, Reading, October 3-7, pp. 43-48, 2005.

[14] Grandi F., Mandreoli F., Martoglia R., Ronchetti E., Scalas M.R., Tiberio P., Semantic Web Techniques for Personalization of eGovernment Services. In Proceedings of ER SemWAT 2006, LNCS 4231, pp. 435-444, 2006.

[15] Groves R.M., Couper M.P., Nonresponse in Household Surveys. New York, Wiley, 1998.

[16] IAP2, IAP2 Public Participation Spectrum, Available at: http://www.iap2.org/associations/4748/files/spectrum.pdf

[17] Kobsa A., Koenemann J., Pohl W., Personalized hypermedia presentation techniques for improving online customer relationships. The Knowledge Engineering Review, 16(2):111–155, 2001.

[18] Lieberman H. L., An agent that assists web browsing. In Proceedings of IJCAI 95, pp. 924–929, San Mateo, CA, USA. Morgan Kaufmann Publishers Inc., 1995.

[19] Magoutas B., Chalaris C., Mentzas G., A Semantically Adaptive Interface for Measuring Portal Quality in e-Government. In Intelligent User Interfaces: Adaptation and Personalization Systems and Technologies, Hershey, PA: IGI Global, 2008.

[20] Montgomery A.L., Smith M.D., Prospects for Personalization on the Internet. In Journal of Interactive Marketing Volume 23, Issue 2, Pages 130-137, 2009.

[21] Nasraoui O., World Wide Web Personalization, In J. Wang (ed), Encyclopedia of Data Mining and Data Warehousing, Idea Group, 2005.

[22] OECD, Promises and Problems of EDemocracy: Challenges of online citizen engagement. Paris: OECD, 2003.

[23] Pazzani M., Billsus D., Content-Based Recommendation Systems. In P. Brusilovsky, A. Kobsa, & W. Nejdl, The Adaptive Web: Methods and Strategies of Web Personalization, Berlin Heidelberg NewYork: Springer-Verlag, 2007.

[24] Perkowitz M., Etzioni O., Adaptive web sites: Automatically synthesizing web pages. In Proceedings of AAAI-98, Madison, WI, USA, 1998.

[25] Pieterson W., Ebbers W., van Dijk J., Personalization in the public sector: an inventory of organizational and user obstacles towards personalization of electronic services in the public sector, *Government Information Quarterly*, Vol. 24 No.1, pp.148-64, 2007.

[26] Schafer J.B., Frankowski D., Herlocker J., Sen S., Collaborative filtering recommender systems. In: Brusilovsky, P., Kobsa, A., Neidl, W. (eds.): The Adaptive Web: Methods and Strategies of Web Personalization. Lecture Notes in Computer Science, Vol. 4321. Springer-Verlag, Berlin Heidelberg New York 291-324, 2007.

[27] Stojanovic N., Stojanovic L., Hinkelmann K., Mentzas G., Abecker A., Fostering self-adaptive e-government service improvement using semantic technologies. AAAI Spring Symposium: The Semantic Web meets eGovernment, Stanford University, California, USA, March 27-29, 2006.

[28] Wang F. H., Shao H. M., Effective personalized recommendation based on time-framed navigation clustering and association mining. Expert Systems with Applications, 27(3), 365–377, 2004.

[29] Wang Y., Kobsa A., Technical Solutions for Privacy-Enhanced Personalization. In C. Mourlas and P. Germanakos (eds.): Intelligent User Interfaces: Adaptation and Personalization Systems and Technologies. pp 326-353, Hershey, PA: IGI Global, 2009.

[30] Won K., Personalization: Definition, Status, and Challenges Ahead. Journal of Object Technology, 1(1): pp. 29-40, 2002.